\documentclass[runningheads]{llncs}
\UseRawInputEncoding

\usepackage{graphicx}
\usepackage{array,multirow}
\usepackage[T1]{fontenc}

\usepackage{dirtytalk}

\usepackage{amssymb}
\usepackage{mathtools}
\usepackage{dsfont}

\newcommand{\R}[0]{\mathds{R}} 
\newcommand{\C}[0]{\mathds{C}} 
\providecommand{\mb}[1]{\mathbf{#1}}
\providecommand{\mbx}{\mb{x}}
\providecommand{\mby}{\mb{y}}
\providecommand{\mbs}{\mb{s}}

\usepackage{graphicx}

\begin{document}
\title{Undersampled MRI Reconstruction with Side Information-Guided Normalisation}
\titlerunning{Undersampled MRI Reconstruction with SIGN}

\author{Xinwen Liu\inst{1} \and
Jing Wang\inst{2} \and Cheng Peng\inst{3} \and Shekhar S. Chandra \inst{1} \and Feng Liu\inst{1} \and S. Kevin Zhou\inst{4,5}}

\authorrunning{X.Liu, et al.}
\institute{School of Information Technology and Electrical Engineering, \\The University of Queensland, Brisbane, Australia.\\
\and The Commonwealth Scientific and Industrial Research Organisation, \\ Canberra, Australia. \\
\and Johns Hopkins University, Baltimore, USA. \\
\and School of Biomedical Engineering \& Suzhou Institute for Advanced Research
Center for Medical Imaging, Robotics, and Analytic Computing \& LEarning (MIRACLE), 
University of Science and Technology of China, Suzhou, 215123, China.\\
\and Key Lab of Intelligent Information Processing of Chinese Academy of Sciences (CAS), Institute of Computing Technology, CAS, Beijing, 100190, China.}
\maketitle
\begin{abstract}
    Magnetic resonance (MR) images exhibit various contrasts and appearances based on factors such as different acquisition protocols, views, manufacturers, scanning parameters, etc. This generally accessible appearance-related side information affects deep learning-based undersampled magnetic resonance imaging (MRI) reconstruction frameworks, but has been overlooked in the majority of current works. In this paper, we investigate the use of such side information as normalisation parameters in a convolutional neural network (CNN) to improve undersampled MRI reconstruction. Specifically, a \textbf{S}ide \textbf{I}nformation-\textbf{G}uided \textbf{N}ormalisation (SIGN) module, containing only few layers, is proposed to efficiently encode the side information and output the normalisation parameters. We examine the effectiveness of such a module on two popular reconstruction architectures, D5C5 and OUCR. The experimental results on both brain and knee images under various acceleration rates demonstrate that the proposed method improves on its corresponding baseline architectures with a significant margin.

\keywords{Deep Learning \and Undersampled MRI Reconstruction}
\end{abstract}

\section{Introduction}
Magnetic resonance imaging (MRI) is a non-invasive imaging modality that produces high contrast in vivo imaging of soft-tissue with non-ionizing radiation. However, due to hardware constraints, MRI suffers from a long scanning time, which impedes its application to real-time imaging. Undersampling the measurements is a common approach that accelerates the imaging process, but can result in blurriness and artifacts that are not suitable for diagnosis purposes. 

Deep learning (DL) has been widely studied to reconstruct the high-quality images from the undersampled measurements \cite{chandra2021deep,guo2021over,knoll2020advancing,recht2020using,zhou2019handbook}. Initially, convolutional neural networks (CNN) have been employed \cite{han2019k,wang2016accelerating,yang2017dagan} to directly map the undersampled images or measurements to fully-sampled ones. Later, imaging model has been integrated into the learning pipeline, and model-based learning has achieved the state-of-the-art performance \cite{aggarwal2018modl,hammernik2018learning,knoll2019assessment,liu2021regularization,qin2018convolutional,schlemper2017deep,sriram2020end,sriram2020grappanet,yang2017dagan,yang2016deep,zhou2020dudornet}. More recently, transformers have been integrated into the CNN-based networks for MRI undersampled reconstruction \cite{feng2021task,guo2022reconformer,korkmaz2022unsupervised,zhou2022dsformer}. These networks can reconstruct MR images with a high fidelity.

\begin{figure}[t]
\begin{center}
\includegraphics[width=\textwidth]{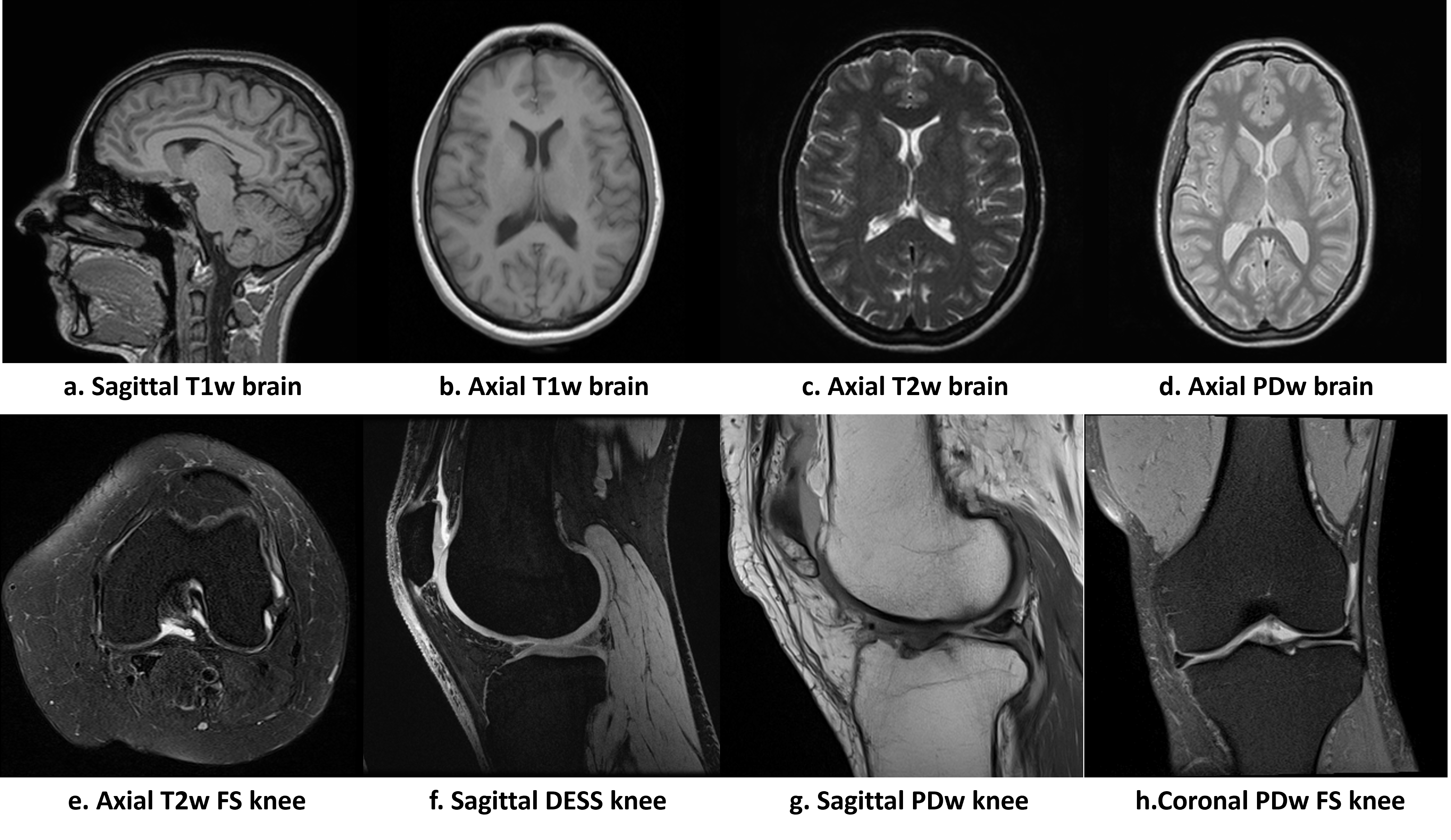}
\caption{MR images of various views, anatomies and acquisition protocols acquired on different machines. a and d are from IXI database; b, c, e, g and h are from fastMRI database \cite{knoll2020fastmri,zbontar2018fastmri}; and f is from SKM-TEA database \cite{desai2021skm}.} \label{fig1}
\end{center}
\end{figure}

MR images present various appearances based on different acquisition protocols, imaging views, scanning parameters, manufacturers, etc. Fig.~\ref{fig1} shows examples of MR images with diverse appearances. When comparing b,c and d, we can observe that for axial brain images, varying acquisition protocols results in differences in brightness, contrast and saturation. This image style variations with acquisition protocols can also be observed by comparing f and g. In addition, due to the hardware differences, images acquired on different machines with divergent scanning parameters present variant noise levels and resolutions. The same subject with different views also shows dissimilar structures. All these differences during acquisitions result in different image styles, which can affect the performance of undersampled reconstruction algorithms \cite{guo2021multi,liu2021universal,wei2020tuning,zhou2020review}. 

Based on such an observation and inspired by recent works on style-related normalisation \cite{gu2021adain,huang2017arbitrary,khan2021switchable,liu2021universal}, we propose a \textbf{S}ide \textbf{I}nformation-\textbf{G}uided \textbf{N}ormalisation (SIGN) module that can be inserted in CNN-based reconstruction networks to obtain higher quality reconstructed image. Specifically, the SIGN module is a sub-network that contains a few linear, normalisation, and non-linear layers. We leverage the side information, which is commonly accessible, and encode it in the SIGN module. SIGN's outputs are used to normalise the feature maps generated by the reconstruction backbone. In this study, we inserted the SIGN modules into the popular D5C5 \cite{schlemper2017deep} and OUCR \cite{guo2021over} backbones. The experiments on brain and knee images under $4\times$ and $6\times$ acceleration show that these networks can obtain significantly better performances by including SIGN modules. 

\section{Methods}
\subsection{Problem formulation}
Reconstructing an image $\mbx \in \C^N$ from undersampled measurement  $\mby \in \C^M$ ($M \ll N$) with side information is formulated as an inverse problem \cite{schlemper2017deep,zhou2019handbook}: 

\begin{equation}
  \label{sparse_coding}
\begin{aligned}
& \underset{\mbx}{\text{min}}
& & \mathcal{R}(\mbx, \mbs) + \lambda \| \mby - \mb{F}_u \mbx \|^2_2,
\end{aligned}
\end{equation}
where $\mbs= [\mbs_e, \mbs_c]$ represents the side information encoded as input vectors to the model, $\mathcal{R}$ is the regularisation term on $\mbx$ and $\mbs$, $\lambda \in \R$ denotes the balancing coefficient between $\mathcal{R}$ and data consistency (DC) term, $ \mb{F}_u \in \C^{M\times N}$ is an undersampled Fourier encoding matrix. In model-based learning, Eq.~\eqref{sparse_coding} is incorporated in the network architecture with $\mathcal{R}$ approximated by the convolution and linear layers.

We divide the side information into two types: categorical variables $\mbs_e = (s_{e}^1,s_{e}^2, ..., s_{e}^{n_{1}})$ and continuous variables $\mbs_c = (s_{c}^1,s_{c}^2, ..., s_{c}^{n_{2}})$ based on the specific piece of information. For attributes such as views, acquisition protocols and manufacturers, they are in distinct groups with a finite number of categories. We use embedding vectors to represent each of these categorical variables. The scanning parameters, such as repetition time (TR), echo time (TR) and flip angles, are continuous variables, which are stacked as vectors. Formally, for each categorical variable $s_{e}^i, i = 1,2,...,n_{1}$, we can obtain the embedded vector $\mb{V}_{e}^i$ as:
\begin{equation}
    \mb{V}_{e}^i = Embedding(s_{e}^i).
\end{equation}

\noindent For the continuous variables $\mbs_c$, we have:
\begin{equation}
\mb{V}_{c} = \mb{W}\mbs_c + \mb{b},
\end{equation}
where $\mb{V}_{c}$ is the representation of continuous information, and $\mb{W}$ and $\mb{b}$ are parameters of the linear mapping layer.

\subsection{Side Information-Guided Normalisation (SIGN) module}
\begin{figure}[h]
\begin{center}
\includegraphics[width=\textwidth]{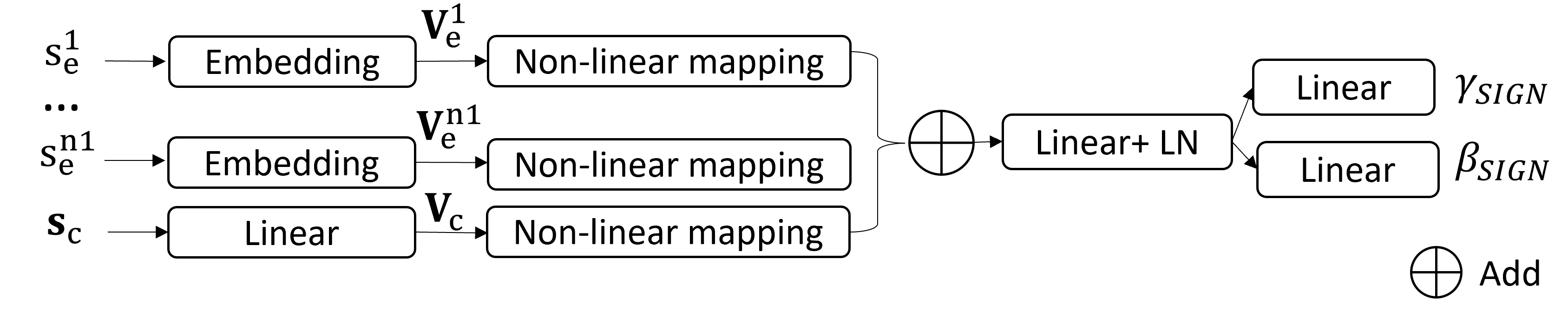}
\caption{The architecture of Side Information-Guided Normalisation (SIGN) module. The inputs are the side information, and the outputs are the parameters to interact with the reconstruction backbone.} \label{fig2}
\end{center}
\end{figure}

To integrate the side information into a network and interact with the main reconstruction network, we propose a Side Information-Guided Normalisation (SIGN) module. Fig.~\ref{fig2} shows the architecture of the proposed SIGN module, which contains linear layers, non-linear activation function (ReLU) and layer normalisation (LN). The inputs to the SIGN module are the side information and the outputs are the parameters to be fed into the reconstruction backbone. 

First, the side information $\mbs$ is encoded into vectors $\{\mb{V}_{e}^i\}$ and $\mb{V}_{c}$. Then, these vectors are passed through the non-linear mapping blocks in parallel. Each of these blocks comprise a fully-connected layer, LN, and ReLU. The latent individual side information features are merged together by element-wise additive operation. At the end, the merged features are fed into another fully-connected layers to further model relationships among each piece of side information. The output parameters ${\gamma}_{SIGN}$ and ${\beta}_{SIGN}$ are sent into the reconstruction backbone.  

Motivated by \cite{gu2021adain,huang2017arbitrary,khan2021switchable,liu2021universal}, the changes of contrasts and texture distributions in images can be mainly described by the mean and variance of feature maps in a deep CNN. Thus, we propagate the encoded side information to control the mean and standard deviation of each feature map in the reconstruction backbone. Note before applying the affine parameters on each layer, we normalize the feature maps per instance to a zero mean and a unit standard deviation. Since different layers have different feature distributions, the SIGN module is not shared cross layers. Formally, for layer $l$ with feature map $h_{l}$ in the reconstruction backbone:

\begin{equation}
\textrm{SIGN}(h_{l})= {\gamma}_{SIGN}^{l}\left(\frac{h_{l}-\mu(h_{l})}{\sigma(h_{l})}\right)+{\beta}_{SIGN}^{l},
\end{equation}
where ${\gamma}_{SIGN}^{l}$ and ${\beta}_{SIGN}^{l}$ are the outputs of the SIGN module, and $\mu(h_{l})$ and $\sigma(h_{l})$ are the mean and standard deviation of the feature map, respectively, computed across spatial dimensions ($H\times W$) independently for each channel.

\subsection{Reconstruction with D5C5 and OUCR backbones}
\begin{figure}[]
\begin{center}
\includegraphics[width=0.8\textwidth]{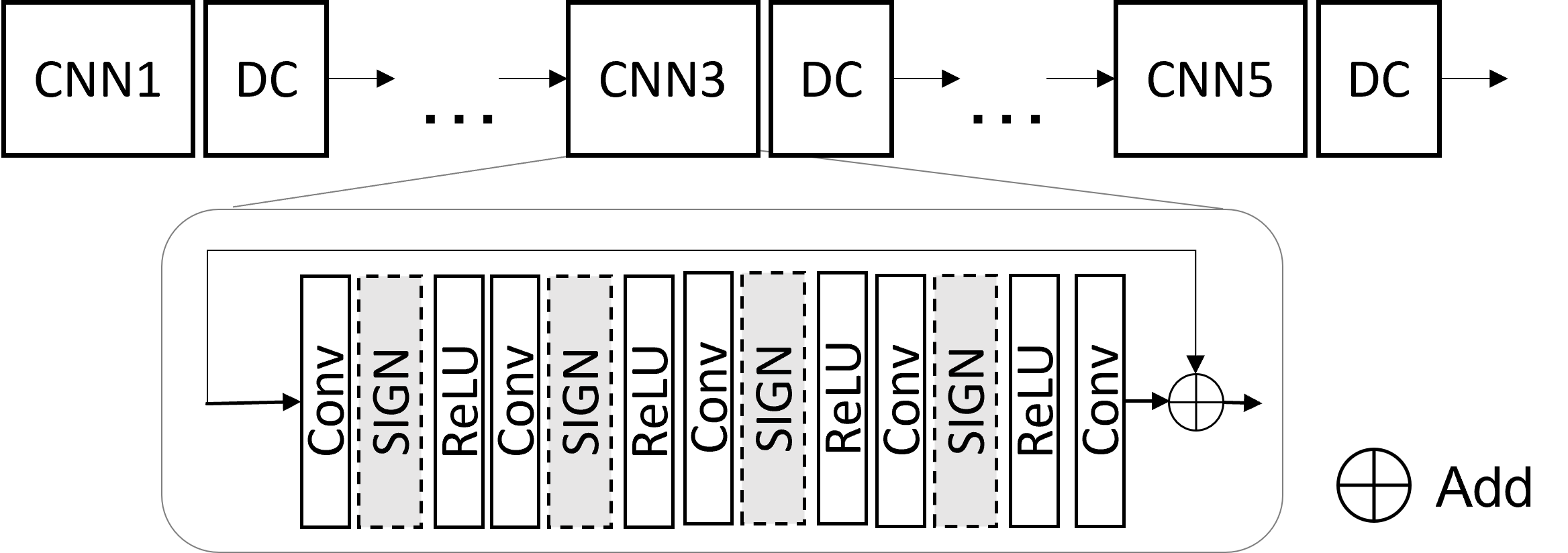}
\caption{The D5C5 reconstruction architecture with the SIGN modules inserted after convolution layers.} \label{fig3}
\end{center}
\end{figure}

D5C5 \cite{schlemper2017deep} and OUCR \cite{guo2021over} are high performance backbones for undersampled MRI reconstruction. D5C5 is a deep cascaded network that contains five interleaved CNN blocks and data consistency (DC) layers. In each of the CNN block, there are five convolutional layers with ReLU activation function and a residual connection at the end. As shown in Fig.~\ref{fig3}, we insert the SIGN modules into the CNN blocks after each of the convolutional layers, except the last one. 

\begin{figure}[h]
\begin{center}
\includegraphics[width=\textwidth]{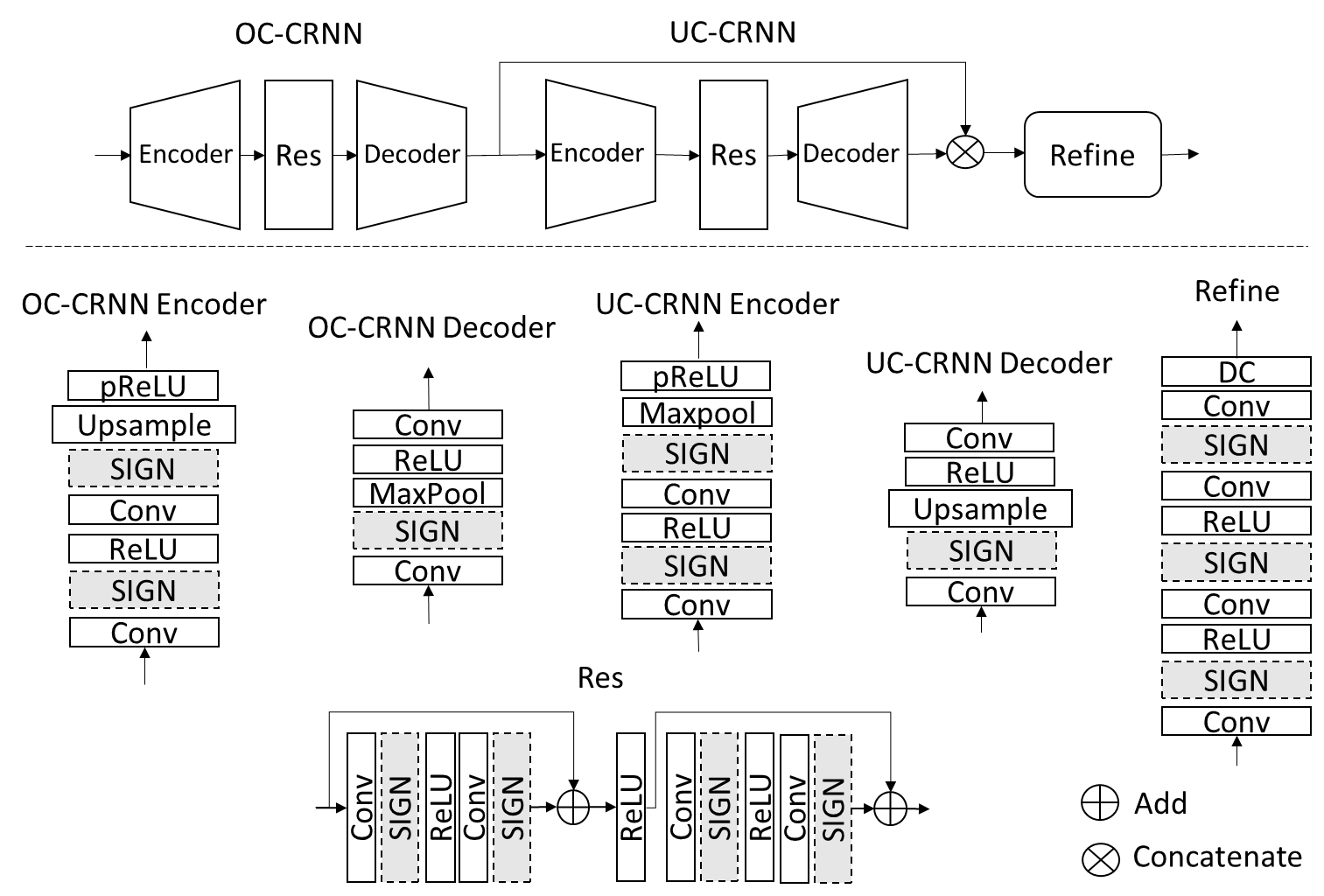}
\caption{The OUCR reconstruction architecture with the SIGN model inserted after convolution layers.} \label{fig4}
\end{center}
\end{figure}

OUCR is a convolutional recurrent neural network (CRNN) that contains the over complete CRNN (OC-CRNN), the under complete CRNN (UC-CRNN) and a refine module, as shown in Fig.~\ref{fig4}. The encoder in UC-CRNN uses maxpooling to reduce the feature map size before being fed into the residual block and the decoder which upsamples the feature map to the original size. The OC-CRNN enlarges the feature map size in the encoder with the upsampling operations. The enlarged feature maps are fed into the residual block and the decoder where the feature maps is down-sampled to the original size. Both OC-CNN and UC-CNN contain an ordinary residual block of convolutions with a skip connection. At the end of the network, a refine block with stacked convolutional layers and ReLU activation functions refine and output the reconstructed results. More details of OUCR can be found in \cite{guo2021over}. The SIGN modules are inserted after each convolutional layer, except the last convolutional layers in the decoder and refine module, which are used to reserve varied feature maps. The numbers of hidden nodes in the fully-connected layers of SIGN modules are assigned to be the same as the number of feature maps in the corresponding layers, except the one in the linear layer after add operation, which is double of the feature map numbers. 

The networks are trained with a mean-absolute-error loss. First, we pretrain the backbone and SIGN modules together. To better optimise the networks, we fix the parameters of the convolutional layers, and only fine-tune the parameters of the linear layers in SIGN modules. 

\section{Experiments and results}
\subsection{Datasets and network configuration}
We evaluate the proposed method on both brain and knee datasets. The brain dataset contains DICOM images from the NYU fastMRI \footnote{fastmri.med.nyu.edu} database \cite{knoll2020fastmri,zbontar2018fastmri} and IXI database\footnote{https://brain-development.org/ixi-dataset/}. The side information in the brain dataset include contrast (T1-weighted, T2-weighted, proton-density-weighted, and magnetic resonance angiogram), views (axial and sagittal), source (Siemens, Philip and GE), and scanning parameters (repetition time, echo time and flip angle). The knee dataset is comprised of DICOM images from the NYU fastMRI database and SKM-TEA database\cite{desai2021skm}. The side information in the knee dataset contains contrast (T2-weighted fat saturated,proton-density-weighted, proton-density-weighted fat saturated, and double echo steady state) and views (axial, sagittal and coronal) and scanning parameters (repetition time, echo time and flip angle). 
It is noted that for IXI database, the scanning parameters are unknown, so we code it as zeros, and the images are indistinguishable from the manufacturers, so we code it as unknown source. In total, there are 10,534 brain slices from 437 volumes and 11,406 knee slices from 246 volumes. For each dataset, 80\% of the volumes are randomly extracted for training, 10\% for validation, and 10\% for testing.

For this experiment, all the data are emulated single-coil magnitude-only images. We crop and resize the images into $320\times320$ and retrospectively undersampled the Cartesian k-space with 1D Gaussian sampling pattern. The evaluation is conducted under acceleration rates of 4$\times$ and 6$\times$ using structural similarity index (SSIM) and peak signal-to-noise ratio (PSNR) as performance metrics. We calculate and report the mean value of the metrics for reconstructed test images.

The experiments are carried out on NVIDIA SXM-2 Tesla V100 Accelerator Units. The networks are implemented on Pytorch and trained using the Adam optimizer with a weight decay to prevent over-fitting.

\subsection{Results}
\label{results_section}
The quantitative comparison of the proposed method and baseline models is shown in Table~\ref{tab1}. The first row shows the metrics of the undersampled images reconstructed directly from the zero-filled $k$-space. D5C5 (row 2) and OUCR (row 4) both improve the PSNR and SSIM over the undersampled images, with OUCR outperforming D5C5 on all cases, which is consistent with \cite{guo2021over}. Comparing D5C5 and D5C5$+$SIGN, the proposed design improves PSNR by about 2$-$2.5dB on brain images, and 1.5dB on knee images. The improvements are also observed on SSIM, with about 1.5$\%$ and 2$\%$ on brain and knee images, respectively. For the experiments on OUCR backbone, the performances of OUCR$+$SIGN are consistently better than those of OUCR. Overall, the experiments show that the simple and effective design of SIGN improves the baseline model with a considerable margin. OUCR$+$SIGN performs slightly lower than D5C5+SIGN. The reason could be due to the complexity of OUCR. Different from D5C5 which has a unified straightforward architecture, OUCR has recurrent designs, multiple residual connections, and varying sizes of feature maps~\cite{guo2021over}. More careful consideration in inserting SIGN module has the potential to improve the results but is out of the scope of this paper.

\begin{table}[h]
\centering
\caption{Quantitative comparison on the reconstruction of brain and knee images under acceleration rates of $4\times$ and $6\times$.}\label{tab1}

\begin{tabular}{r|c|c|c|c|c|c|c|c}
\hline
\multirow{3}{*}{} & \multicolumn{4}{c|}{$4\times$} & \multicolumn{4}{c}{$6\times$} \\ \cline{2-9}

\multicolumn{1}{c|}{} & \multicolumn{2}{c|}{PSNR(dB)} & \multicolumn{2}{c|}{SSIM($\%$)} & \multicolumn{2}{c|}{PSNR(dB)} & \multicolumn{2}{c}{SSIM($\%$)} \\ \cline{2-9}
& Brain & Knee & Brain & Knee & Brain & Knee & Brain & Knee  \\ \hline

Undersampled & 29.92 & 30.04 &  79.27 &  79.16  & 22.03 & 24.63 & 60.02 &  66.49  \\ \hline
D5C5 & 39.29 &  35.34 & 96.89  & 90.69  & 36.03  & 33.23  & 94.62  & 87.51 \\\hline
D5C5+SIGN & \textbf{41.70}  & \textbf{37.18} & \textbf{97.93}  & \textbf{93.02}  &  \textbf{38.16} & \textbf{34.66}  &  \textbf{96.39}   &  \textbf{89.75} \\\hline
OUCR & 40.29 & 36.48 & 97.44  & 92.29 & 36.93  & 34.16  & 95.58  & 89.17  \\\hline
OUCR+SIGN & \textbf{41.64}  & \textbf{37.06} & \textbf{97.94}  & \textbf{92.93} & \textbf{37.96}  & \textbf{34.47} & \textbf{96.27}  & \textbf{89.51}  \\\hline
\end{tabular}
\end{table}

\begin{figure}[h]
\begin{center}
\includegraphics[width=\textwidth]{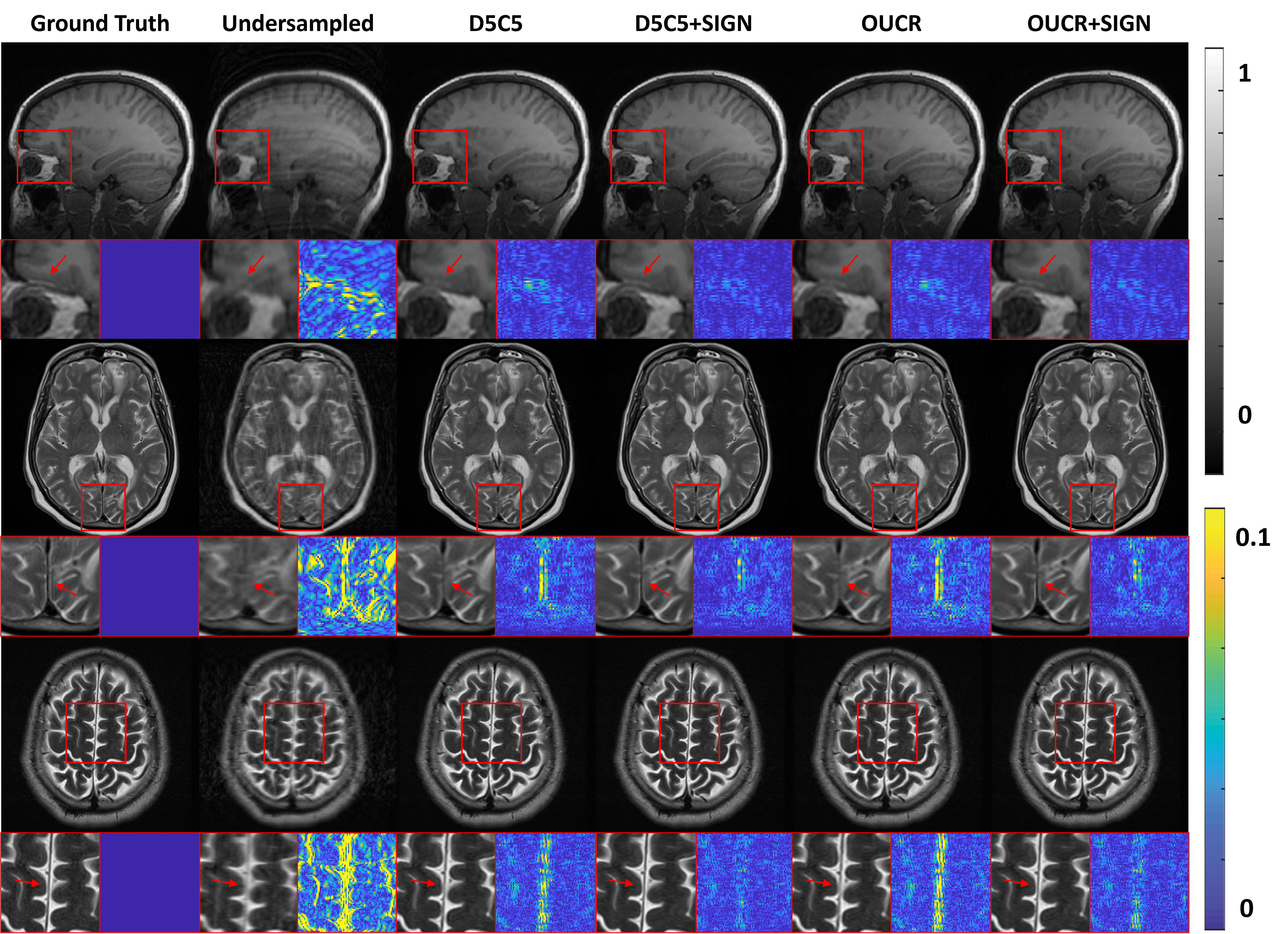}
\caption{The reconstructed brain images under 4$\times$ acceleration with different methods.} \label{fig5}
\end{center}
\end{figure}

Fig.~\ref{fig5} shows the examples of reconstructed brain images under 4$\times$ acceleration with different methods.
The first column is the fully-sampled images and the second column represents the undersampled images, which are blurry and full of artifacts. As shown in the third column, D5C5 can reconstruct the images. In the forth column, D5C5$+$SIGN further improves the images with more detailed information preserved, as pointed by the arrows. Similar trend can also be observed by comparing the fifth column (OUCR) and the sixth column (OUCR$+$SIGN). OUCR with the SIGN inserted in the network recovers the images with higher visual fidelity.  

\subsection{Further analysis}
This section further investigates the effectiveness of the proposed SIGN module. As an example, we test the D5C5$+$SIGN model in recovering $4\times$ accelerated brain images. In the first experiment, we set the side information of each image with a random label. The results show that the PSNR drops from 41.70dB to 40.93dB. Then, we evaluate the reconstruction with wrong side information. For example, for the image with axial view, we assign it to be the sagittal view. The performance drops by approximately 1dB. 

In addition, the importance of each side information is evaluated by masking out one or more branch of the side information in SIGN. When only using the contrast information, the performance goes down to 41.39dB. With view information alone, PSNR is 40.94dB. If both contrast and view information are used, PSNR becomes 41.42dB. More comparison and results can be found in the supplementary material (Table~S1-2). Overall, reconstructing images with wrong or missing side information leads to a worse performance.

\section{Conclusion}
In this paper, we propose to utilise the side information as normalisation parameters to improve the undersampled MRI reconstruction. Experimental results demonstrate the proposed approach can improve the baseline models significantly on both brain and knee images under $4\times$ and $6\times$ acceleration. While the current design is based on the popular D5C5 and OUCR architectures, the proposed framework is extendable to other network architectures. The current study is based on the single-coil magnitude-only images for proof-of-concept. In the future, we will adapt the network to multi-coil complex-valued datasets.

\bibliographystyle{splncs04}
\bibliography{mybibliography}

\newpage
\section{Supplementary Materials}

\begin{table}[h]
\renewcommand{\thetable}{S\arabic{table}}
\centering
\caption{The reconstruction results with true side information, randomly generated side information and wrong side information on $4\times$ accelerated brain images.}\label{tabs1}

\begin{tabular}{c|c|c|c|c|c}
\hline
 \multicolumn{3}{c|}{PSNR(dB)} & \multicolumn{3}{c}{SSIM($\%$)} \\  \hline
True & Random & Wrong & True & Random & Wrong  \\ \hline
\textbf{41.70} & 40.93 & 40.70 & \textbf{97.93} & 97.69 &97.61 \\ \hline

\end{tabular}
\end{table}
\begin{table}[h]

\renewcommand{\thetable}{S\arabic{table}}
\centering
\caption{The impact of side information on the reconstruction of $4\times$ accelerated brain images. c: contrast, v: view, s: source.}\label{tabs2}

\begin{tabular}{c|c|c|c|c|c||c|c|c|c|c|c}
\hline \multicolumn{6}{c||}{1 side information} & \multicolumn{6}{c}{2 side information} \\\hline

\multicolumn{3}{c|}{PSNR(dB)} & \multicolumn{3}{c||}{SSIM($\%$)} & \multicolumn{3}{c|}{PSNR(dB)} & \multicolumn{3}{c}{SSIM($\%$)} \\\hline 
c & v & s & c & v & s & c+v & c+s & v+s & c+v & c+s & v+s  \\ \hline
41.39 & 40.94 & 41.20 & 97.86 & 97.71 & 97.76 &41.42 & 41.67 & 41.25 & 97.87 & 97.91 &97.78 \\\hline
\end{tabular}
\end{table}

\end{document}